\documentclass[prd,aps,showpacs,floats,letterpaper,floatfix,groupedaddress]{revtex4}
\usepackage{amssymb,amsmath}
\usepackage{graphicx,epsf, epsfig, amssymb}

\usepackage{dcolumn,epsfig}

\def\be{\begin{equation}}
\def\ee{\end{equation}}
\def\beq{\begin{eqnarray}}
\def\eeq{\end{eqnarray}}

\makeatletter
\def\equalsfill{$\m@th\mathord=\mkern-7mu
\cleaders\hbox{$\!\mathord=\!$}\hfill
\mkern-7mu\mathord=$}
\makeatother

\def\defn{\stackrel{\text{def}}{\hbox{\equalsfill}}}

\begin{document}

\title{Gravitational Larmor formula in higher dimensions}

\author{Vitor Cardoso}
\email{vcardoso@phy.olemiss.edu} \affiliation{Department of Physics and Astronomy, The University of
Mississippi, University, MS 38677-1848, USA}

\author{Marco Cavagli\`a} \email{cavaglia@phy.olemiss.edu}
\affiliation{Department of Physics and Astronomy, The University of
Mississippi, University, MS 38677-1848, USA}

\author{Jun-Qi Guo} \email{jguo@phy.olemiss.edu}
\affiliation{Department of Physics and Astronomy, The University of
Mississippi, University, MS 38677-1848, USA}

\date{\today}

\begin{abstract}
The Larmor formula for scalar and gravitational radiation from a pointlike particle is derived in any even
higher-dimensional flat spacetime. General expressions for the field in the wave zone and the energy flux are
obtained in closed form. The explicit results in four and six dimensions are used to illustrate the effect of
extra dimensions on linear and uniform circular motion. Prospects for detection of bulk gravitational
radiation are briefly discussed.
\end{abstract}


\maketitle
\section{Introduction}
\label{intro}
The interest in higher-dimensional scenarios has increased ever since the first attempts by Kaluza and Klein
to unify gravity and electromagnetism \cite{kaluzaklein}. Recent proposals include large extra-dimensional
models \cite{hamed} and braneworlds \cite{Randall,Maartens}, which have been advocated as a possible solution
to the hierarchy problem of gauge couplings. Moreover, higher-dimensional theories generally possess more
degrees of freedom, thus providing a richer arena to describe physical phenomena.

Higher-dimensional models of gravity generally exhibit two potentially testable characteristics: (i) Newton's
force at short distances no longer scales as $r^{-2}$ \cite{hamed, Randall, Maartens, garrigatanaka} and (ii)
generation and emission of gravitational radiation differ from the four-dimensional analogues, leading to
observable effects \cite{cardosoddim,CCP,radiationhigherd}. Searches for deviations from Newton's
inverse-squared law are currently in progress \cite{Hoyle:2004cw}. Gravitational events in particle colliders
and cosmic ray extensive airshowers \cite{bhprod,ccg} may also provide indirect evidence of large extra
dimensions. The physics of generation and emission of gravitational waves in higher-dimensional scenarios has
hardly been explored at all.

The aim of this paper is to derive an exact formula for the radiation field of a charge moving in a even
higher-dimensional spacetime. Here, charge will stand either for scalar or gravitational charge. There are
several motivations for this study. The most popular models of extra dimensions allow for gravitational and
scalar degrees of freedom in the bulk (e.g.\ the radion \cite{Goldberger:1999uk}). Observable effects of bulk
gravitational and scalar radiation on the visible brane could provide a valuable signature of the existence of
extra dimensions. Moreover, most radiation phenomena can be analyzed in flat geometries by means of scalar
fields, provided that careful cutoffs are imposed. (See for instance Mironov and Morozov
\cite{radiationhigherd}.) A full tensorial analysis of Einstein equations in higher dimensions shows indeed
that the gravitational degrees of freedom are either equivalent to a massless scalar equation in flat spacetime
with an appropriate source term \cite{cardosoddim,misner}, or to a massless scalar field plus a massive field
with source term including brane contributions \cite{garrigatanaka, giddingsrandall}. Therefore, scalar fields
can be used as a simple model mimicking the more complex tensor field.

This paper is organized as follows. In Section II the field in the radiation zone and the Larmor formula
are derived. Section III deals with some special cases (linear and uniform circular motion). Conclusions are
presented in Section V. Unless explicitly stated, throughout the paper the speed of light is set equal to
unity.
\section{Computation of the Larmor formula}
\label{sec:theory}
A massless scalar field in $D$ dimensions ($x=0\dots D-1$) with source $S(x)$ satisfies the wave equation
\be
\Box\, \varphi_D=S(x)\,.
\label{KG}
\ee
The source is a minimally-coupled pointlike particle with nonzero mass
\be
S(x)= \frac{\alpha_D}{\sqrt{g(x)}}\int d\tau\ \delta^{(D)}\left[(x^{\mu}-x_p^{\mu}(\tau)\right]\,,
\label{sourceterm}
\ee
where the particle worldline is defined by $x^\mu-x^\mu_p(\tau)=0$, $\tau$ is the proper time along a
geodesic, $g(x)$ is the determinant of the metric, and $\alpha_D$ is the coupling constant. The (retarded)
solution of Eq.\ (\ref{KG}) is
\be
\varphi_{D}(x)=\int d^{D}x'\, \mathcal{G}_{D}(x,x') S(x')\,,
\label{Greensolution}
\ee
where $\mathcal{G}_{D}(x,x')$ is the retarded $D$-dimensional Green function of the wave operator, $\Box\,
\mathcal{G}_{D}(x,x')=\delta^{(D)}(x-x')$. According to the discussion in the introduction, the metric
$g_{\mu\nu}$ is replaced with the Minkowski metric. The Green functions in $D=2$ and $D=3$ are
\be
\mathcal{G}_{2}(z)=\frac{1}{2}\theta (z)\,,
\label{Green2}
\ee
and
\be
\mathcal{G}_{3}(z)=\frac{1}{2\pi}\frac{\theta (z)}{\sqrt{(x^{0}-x^{0}_{p})^{2}-R^{2}}}\,,
\label{Green3}
\ee
respectively. Here $z=x^{0}-x^{0}_{p}(t')-|{\bf x}-{\bf x}_{p}(t')|$, $ R=|{\bf x}-{\bf x}_{p}(t')|$, and
$\theta(z)$ is the Heaviside step function. The Green function in $D\ge 4$ dimensions can be obtained from
the Green function in $D-2$ dimensions through the recursive relation \cite{hassani,SoodakTiersten}:
\be
\mathcal{G}_{D}(z)=\frac{1}{2\pi R}\frac{d}{dz}\,\mathcal{G}_{D-2}(z)\,,~~~~~~D\ge 4\,.
\label{GreenD}
\ee
Equations (\ref{Green2})-(\ref{GreenD}) show that the Green functions for even-dimensional spacetimes ($D>2$)
have support on the past light cone. The Green function for $D=2$ and for odd-dimensional spacetimes have
also support inside the past light cone, because of their dependence on $\theta(z)$. Therefore, Huygens'
principle is not satisfied in these spacetimes \cite{SoodakTiersten}. (This is also the reason for the
appearance of wakes behind a boat sailing on the two-dimensional surface of a lake.) Since the appearance of
wake phenomena in odd dimensions makes the problem very complex to handle \cite{cardosoddim}, the
analysis below will be limited to even-dimensional spacetimes. The leading term of the Green
function in the far zone is
\be
\mathcal{G}_{2k}(z)=\left (-\frac{1}{2\pi R}\frac{\partial}{\partial R}
\right)^{k-1}\mathcal{G}_{2}(z)\,,
\label{Greendominant}
\ee
where $k=D/2$. From Eqs.\ (\ref{Greensolution}) and (\ref{Greendominant}), it follows that the dominant
term is of order $R^{-k+1}$. The Green function in the far zone is
\be
\mathcal{G}_{2k}(z)=\frac{\delta^{(k-1)}(z)}{2(2\pi R)^{k-1}}
+{\cal O}(R^{-k})\,.
\label{Greendominant2}
\ee
The field in the far zone is found by substituting Eqs.\ (\ref{sourceterm}) and (\ref{Greendominant2}) in Eq.\
(\ref{Greensolution}):
\be
\varphi_{2k}(x)=\alpha_{2k}\int_{-\infty}^{+\infty} d\tau
\left[\frac{\delta^{(k-2)}(z)}{2(2\pi R)^{k-1}}+{\cal O}(R^{-k})\right]\,,
\label{phi}
\ee
where $z$ and $R$ are defined as below Eq.\ (\ref{Green3})  with $t'\to\tau$. Integrating by parts, Eq.\
(\ref{phi}) reads
\be
\varphi_{2k}(x)=\frac{\alpha_{2k}}{2(2\pi R)^{k-1}}
\left(\frac{1}{B}\frac{d}{d\tau}\right)^{k-2}\frac{1}{B}+{\cal O}(R^{-k})\,,
\label{g2}
\ee
where
\be
B\defn -\frac{dz}{d\tau}= \gamma (1-{\bf n} \cdot {\bf v})\,.\label{dzdtau}
\ee
The above result can be rewritten in the useful form
\be
\varphi_{2k}(x)=\frac{1}{2\pi RB}\frac{d}{d\tau}\left
(\varphi_{2k-2}\right)+{\cal O}(R^{-k})\,.
\label{recurrence}
\ee
The field in the wave zone can be computed recursively with Eq.\ (\ref{recurrence}) in any even $D\ge 4$
dimension. The Larmor formula can be easily derived from the stress energy-momentum tensor of the field.
In the asymptotic region, the energy flux per unit time (Poynting vector) is
\be
{\bf T}_{2k}=-\dot\varphi_{2k}{\bf \nabla}\varphi_{2k}\,.
\label{Pointing}
\ee
Substitution of Eq.\ (\ref{g2}) in Eq.\ (\ref{Pointing}) yields
\be
{\bf T}_{2k}= \frac{\alpha_{2k}^2}{4(2\pi R)^{2k-2}}\left[
\left(\frac{1}{B}\frac{d}{d\tau}\right )^{k-1}\frac{1}{B}\right]^2
{\bf n}+{\cal O}(R^{-2k+1})\,,
\ee
where $\bf n$ is the unit vector in the direction of ${\bf x}-{{\bf x}_p}$. The power emitted per unit
of solid angle in the direction $\bf n$ is
\be
\frac{dP_{2k}}{d\Omega_{2k-2}}= \frac{\alpha_{2k}^2}{4(2\pi)^{2k-2}}\frac{B}{\gamma}\left[
\left(\frac{1}{B}\frac{d}{d\tau}\right )^{k-1}\frac{1}{B}\right]^2\,.
\label{powerloss}
\ee
Equation (\ref{powerloss}) is an exact expression.
\section{Linear and circular motion}
\label{sec:specialcases}
It is instructive to consider some special cases of Eq.\ (\ref{powerloss}). The power loss in four ($k=2)$
and six ($k=3$) dimensions are
\beq
&&{\frac{dP_4}{d\Omega_2}}= \frac{\alpha_4^2}{16\pi^2}\frac{[{\bf
a}\cdot\left(\gamma^2(1-{\bf v}\cdot{\bf n}){\bf v}-{\bf n}
\right)]^2}{\gamma^2 (1-{\bf v}\cdot{\bf n})^5}\,,
\label{power4d}\\
&&{\frac{dP_6}{d\Omega_4}}=
\frac{\alpha_6^2}{64\pi^4}\frac{[CF-3E^2/F]^2}{\gamma^8 (1-{\bf v}\cdot{\bf
n})^7}\,,
\label{power6d}
\eeq
respectively. In the above equations, $\bf a$ and $\bf v$ are the acceleration and velocity of the particle
and
\be
C=\gamma^4 ({\bf a}\cdot {\bf v})^\cdot-\frac{{\bf
n}\cdot (\gamma^2{\bf a})^\cdot+\gamma^4({\bf n} \cdot {\bf
a})({\bf a} \cdot {\bf v})}{1-{\bf n} \cdot {\bf v}}\,,~~~
E=\gamma^2[\gamma^2{\bf a} \cdot {\bf v}({1-{\bf n} \cdot {\bf
v}})-{\bf n} \cdot {\bf a}]\,,~~~
F=\gamma ({1-{\bf n} \cdot {\bf v}})\,,
\label{powerdef}
\ee
where dot denotes differentiation w.r.t.\ $x_0$. As is expected from simple relativistic considerations,  it
is straightforward to check that there is no radiation for linear uniform motion. This remains true if the
bulk has finite volume and its spatial boundary is flat, such as the simple ADD scenario with a smooth brane.
However, if the latter is inhomogeneous, radiation is generated \cite{CCP}. This phenomenon is analogous to
electromagnetic diffraction of an electric charge moving near a metal grating (Smith-Purcell effect). It can
be understood by replacing the brane with a set of oscillating image charges. The image configuration
is time dependent because of the brane inhomogeneities and diffraction radiation is generated by the
reflection of the boosted static field on the nearby wall.

The total power emitted from a particle in planar uniform circular motion with radius $R_0$ and angular
frequency $\omega$ is:
\beq
&&P_{{\rm circ},4}=\frac{\alpha_4^2}{12\pi}\gamma^4\omega^4R_0^2\,,
\label{power4dcirc}\\
&&P_{{\rm circ},6}=\frac{\alpha_6^2}{120\pi^2}\gamma^8\omega^6R_0^2(1+4\omega^2R_0^2)\,.
\label{power6dcirc}
\eeq
Equations (\ref{power4dcirc}) and (\ref{power6dcirc}) describe scalar synchrotron power loss in four and six
dimensions, respectively. Assuming that the field coupling constant does not vary too much with $D$, the
synchrotron loss in six dimensions is larger than in four dimensions for angular frequencies greater than
$\omega\sim L^{-1}$, where $L$ is the fundamental length scale. Thus a particle radiates more in higher
dimensions. Equations (\ref{power4dcirc}) and (\ref{power6dcirc}) can be used, at least in principle, for
indirect detection of extra dimensions by measuring the increase in the power loss of a particle on the brane
as function of the Lorentz factor $\gamma$. A power loss scaling as $\gamma^8\omega^6$ would signal the
presence of two additional dimensions.

The above results can be translated to the gravitational case by setting $\alpha_{2k}=\sqrt{G_{D}}m\gamma^2$
\cite{misner}, where $m$ is the mass of the particle and $G_D$ is the Newton constant in $D$ dimensions. If a
scenario with $D-4$ large extra dimensions of size $L$ \cite{hamed} is assumed, $G_D$ is related to the
four-dimensional Newton constant $G_4$ by $G_D\sim L^{D-4}G_4$. Restoring the speed of light $c$, the
expressions for the gravitational power loss in four and six dimensions are
\beq
&&P_{{\rm circ},4}\sim\frac{G_4}{12\pi c^3}m^2\gamma^8 \omega^2(\omega R_0)^2\,,
\label{power4dcirc2}\\
&&P_{{\rm circ},6}\sim\frac{G_4}{120\pi^2 c^5}m^2\gamma^{12}\omega^2(L\omega)^2(\omega
R_0)^2(1+4\omega^2R_0^2/c^2)\,.
\label{power6dcirc2}
\eeq
The energy loss for physical systems can be roughly estimated by comparing the above results to the ordinary
four-dimensional synchrotron radiation, $P_{\rm sync}$, which is emitted by a particle with electric charge $e$
\cite{jackson}:
\beq
&&\frac{P_{{\rm circ},4}}{P_{\rm sync}}\sim\frac{G_4 m^2}{e^2}\gamma^4
\sim 10^{-36}\gamma^4\left(\frac{m}{\rm GeV}\right)^2\,,
\label{power4sync}\\
&&\frac{P_{{\rm circ},6}}{P_{\rm sync}}\sim\frac{G_4 m^2}{e^2 c^2}\gamma^8(L\omega)^2
\sim 10^{-59}\gamma^8\left(\frac{m}{\rm GeV}\right)^2\left(\frac{L}{\rm mm}\right)^2
\left(\frac{\omega}{\rm Hz}\right)^2\,.
\label{power6sync}
\eeq
For a proton, $m\sim 1$ GeV, thus the gravitational emission becomes comparable to the synchrotron emission
for $\gamma\sim 10^9$ (four dimensions) and $\gamma\sim 10^8\,(\omega/{\rm Hz})^{-2}$ (six dimensions). Thus
the gravitational emission is negligible in any current or near-future Earth-based experiment. For instance,
the Large Hadron Collider at CERN will collide protons with $\gamma \sim 10^4$ at a frequency of $\omega \sim
10^4~{\rm Hz}$ \cite{lhc}, which yields $P_{{\rm circ},4}\sim 10^{-20}P_{\rm sync}$ and $P_{{\rm circ},6}\sim
10^{-20}(L/{\rm mm})^2P_{\rm sync}$. However, gravitational emission could become relevant in astrophysical
processes. Magnetic fields larger than $10^{12} {\rm G}$ are thought to occur in neutron stars, active
galactic nuclei and other sources \cite{reviewolinto}. This implies very high frequencies and large $\gamma$
factors. Therefore, indirect detection of extra dimensions by gravitational synchrotron radiation
could be possible.
\section{Conclusions}
\label{sec:conclusions}
We derived a simple expression for scalar and gravitational radiation by point particles in a generic (even)
number of spacetime dimensions. The power loss in scalar and gravitational waves becomes significant for high
frequencies and large Lorentz factors. The enhancement of radiation in extra-dimensional scenarios may lead to
detectable astrophysical effects, such as higher radiation damping around neutron stars and active galactic
nuclei. These results are limited to flat spacetimes. It would be interesting to consider gravitational
radiation in warped scenarios, including possible Kaluza-Klein mode excitation.

\end{document}